\documentclass[letter,fleqn,usenatbib]{mnras}

\usepackage{mathptmx}

\usepackage[T1]{fontenc}
\usepackage{ae,aecompl}

\usepackage{graphicx}	
\usepackage{amsmath}	
\usepackage{amssymb}	

\newcommand\vdelta{{\boldsymbol\delta}}
\newcommand\AU{{\rm AU}}

\title[Eccentricity distribution in the main asteroid belt]{Eccentricity distribution in the main asteroid belt}

\author[Renu Malhotra and Xianyu Wang]{
Renu Malhotra$^{1}$\thanks{E-mail: renu@lpl.arizona.edu}
and Xianyu Wang$^{1,2}$
\\
$^{1}$Lunar and Planetary Laboratory, The University of Arizona, Tucson, AZ 85721, USA\\
$^{2}$School of Aerospace Engineering, Tsinghua University, Beijing, China
}

\date{Accepted XXX. Received YYY; in original form ZZZ}

\pubyear{2016}

\begin{document}
\label{firstpage}
\pagerange{\pageref{firstpage}--\pageref{lastpage}}
\maketitle

\begin{abstract}
The observationally complete sample of the main belt asteroids now spans more than two orders of magnitude in size and numbers more than 64,000 (excluding collisional family members). We undertook an analysis of asteroids' eccentricities and their interpretation with simple physical models. We find that Plummer's (1916) conclusion that the asteroids' eccentricities follow a Rayleigh distribution holds for the osculating eccentricities of large asteroids, but the proper eccentricities deviate from a Rayleigh distribution: there is a deficit of eccentricities smaller than $\sim0.1$ and an excess of larger eccentricities.  We further find that the proper eccentricities do not depend significantly on asteroid size but have strong dependence on heliocentric distance: the outer asteroid belt follows a Rayleigh distribution, but the inner belt is strikingly different.  Eccentricities in the inner belt can be modeled as a vector sum of a primordial eccentricity vector of random orientation and magnitude drawn from a Rayleigh distribution of parameter $\sim0.06$, and an excitation of random phase and magnitude  $\sim0.13$. These results imply that a late dynamical excitation of the asteroids occurred, it was independent of asteroid size, it was stronger in the inner belt than in the outer belt. We discuss implications for the primordial asteroid belt and suggest that the observationally complete sample size of main belt asteroids is large enough that more sophisticated model-fitting of the eccentricities is warranted and could serve to test alternative theoretical models of the dynamical excitation history of asteroids and its links to the migration history of the giant planets.
\end{abstract}

\begin{keywords}
minor planets, asteroids -- solar system: formation -- solar system: general
\end{keywords}



\section{Introduction}

A century ago in these pages, \cite{Plummer:1916} published what appears to have been the first statistical assessment of the orbital data of the minor planets of the solar system. He noted the surprising lack of very small eccentricities amongst the 809 minor planets then known, and he empirically determined that the eccentricity distribution was close to 
\begin{equation}
dN(e) =  C \, e \exp(-\alpha\, e^2) de,
\end{equation}
with $C=59100,\, \alpha=36.5$.  Although Plummer did not call it so, this is a Rayleigh distribution with parameter $\sigma=(2\alpha)^{-{1\over2}}=0.117$.  
In the decades since, the known population of minor planets increased by more than two orders of magnitude, and this has allowed many interesting studies of their physical and dynamical properties. 
Surprisingly, the topic of the eccentricity distribution of observed asteroids, its relation to (and deviations from) the Rayleigh distribution has scarcely been revisited in the past century; only \cite{Beck:1981} appears to have examined it when the sample size had increased by a factor of less than three.  
In a recent paper, \citet{Michtchenko:2016} presented a wide-ranging analysis of the modern data of the main belt asteroids, including their orbital parameters and their physical properties, but did not address the topic of the putative Rayleigh distribution of asteroids' eccentricities.
In the present work, we attempt to remedy this gap. 

Our work is also motivated by the many recent models of an instability-driven dynamical history of the solar system~\citep[see][for a review]{Dones:2015}. Early dynamical events could have left an imprint on the orbital distribution in the asteroid belt, therefore rigorous statistical analysis of the asteroids' eccentricities may provide one possible test of alternative models of the solar system's past evolution.

The large increase in the known population of asteroids and increased knowledge of their dynamics means that there are greater complexities in analyzing the data.  These include defining the boundaries of the main asteroid belt, defining an observationally complete sample, accounting for collisional families of asteroids, and distinguishing between osculating and proper elements.  It is also interesting to examine the size dependence and heliocentric distance dependence of the eccentricity distribution, made possible by the large numbers of asteroids now known.
 In Section~\ref{s:data} we describe the asteroid data samples of interest.  In Section~\ref{s:analysis}, we describe the results of our statistical analysis of the eccentricities and their interpretation with simple physical models.  We summarize and conclude in Section~\ref{s:summary}.

\section{Asteroid data}\label{s:data}

Our primary source of asteroid data is the service known as the ``Asteroids Dynamic Site'', or AstDys-2 (URL http://hamilton.dm.unipi.it/astdys2).
This service (updated daily) provides tables of the orbital elements and the absolute magnitudes of all numbered minor planets, as well as the family membership information of each object.
For each object, three types of orbital elements are provided: osculating elements, analytical proper elements and synthetic proper elements.
The osculating elements describe the heliocentric osculating Keplerian orbit, whereas the proper elements are computed from the osculating elements by removing the perturbing effects of the eight planets, Mercury--Neptune~\citep{Knezevic:2002}.
The analytical proper elements are based on an analytical secular theory of degree 4 in the eccentricities and inclinations and degree 2 in planetary masses; this theory fails for some asteroids, usually those near mean motion resonances with the planets or those with high eccentricities and/or inclinations~\citep{Milani:1994}. The synthetic proper elements, which consist of the three parameters, $(a,e,\sin i)$, are based on a numerical procedure which removes the planetary perturbations more accurately~\citep{Knezevic:2000,Knezevic:2003}.

\begin{figure}
\centering
\includegraphics[scale=0.33,angle=270]{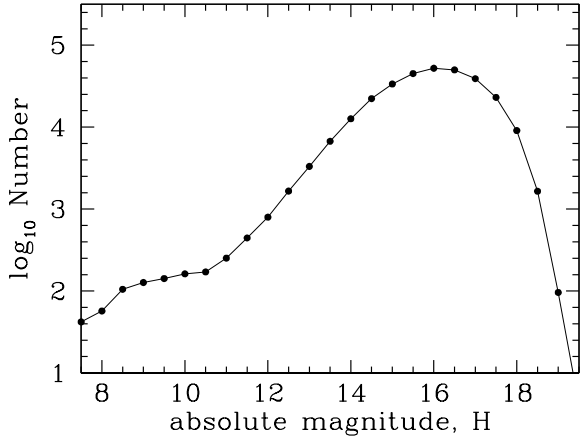}
\caption{Distribution of absolute magnitude, $H$, of main belt asteroids (excluding members of collisional families).
The bin size is $\Delta H=0.5$ and the locations of the bins are indicated by the dots.
}
\label{f:figlHm}
\end{figure}

In the present work, we will use primarily the synthetic proper elements. The proper elements being free of the variations imposed by planetary perturbations are thought to best hold memory of the formation and dynamical history of asteroids. These elements have thus far been employed primarily for the identification and study of collisional families of asteroids, but here we will be focussed on the main asteroid belt as a whole.

At the time of this work, 
the AstDys-2 provided synthetic elements for 406,251 numbered asteroids, and identified 102,147 as members of various asteroid families (as of October 20, 2015).
We adopt the boundaries of the main asteroid belt as $a(1-e)>1.6$~AU and $a<3.8$~AU.  This definition, which is similar to one adopted in many previous studies, excludes asteroids with perihelion distance smaller than Mars' aphelion as well as the Hilda group and the Trojan asteroids (in 3:2 and 1:1 mean motion resonance with Jupiter, respectively). 
For the purpose of the present work, it is prudent to exclude members of collisional families when examining the overall eccentricity distribution of the main belt asteroids (MBAs, henceforth). 
Thus pared, the sample consists of 302,698 numbered asteroids. 
We plot in Fig.~\ref{f:figlHm} the distribution of the absolute magnitude, $H$, of this sample of MBAs. 
We can discern from this plot that these MBAs are observationally incomplete for $H\gtrsim15$ (equivalent diameter $D\lesssim5$~km).  We also note significant changes in the slope of the distribution function near $H\approx8.5$ and $H\approx11$.

\begin{figure}
\centering
\includegraphics[scale=0.33,angle=270]{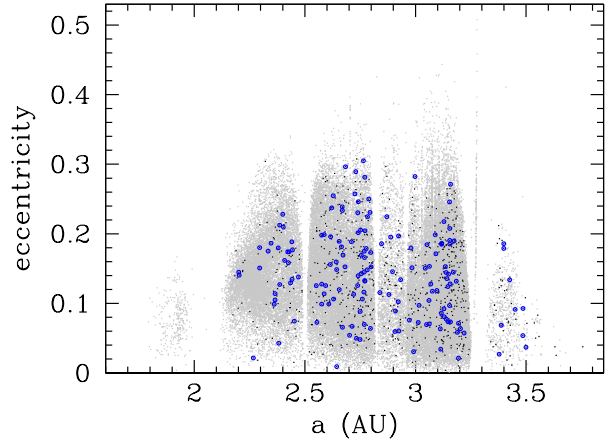}
\caption{Proper eccentricity versus proper semimajor axis of main belt asteroids (excluding members of collisional families).  The observationally nearly complete set is represented in the small grey dots;  the ``nearly primordial'' asteroids, with $H\leq10.8$, are in black, and the ``even more primordial'' asteroids, with $H\leq8.2$, are the blue circles.}
\label{f:figpae}
\end{figure}

We adopt $H\leq15$ for the observationally nearly complete set; there are 64,377 asteroids in this set\footnote{Adopting a slightly different $H$ cut-off, $H\leq14$, yields a smaller sample size but does not substantially change the results reported here.}. 
For context, we provide a scatter plot of $(a,e)$ of these MBAs in Fig.~\ref{f:figpae}.  Visual inspection shows that near-zero eccentricities are rare, particularly in the inner asteroid belt.   This is qualitatively consistent with Plummer's observations of a century ago.

We will also be interested in several subsets of the observationally complete set.  We define two subsets based on size, defined by the changes in slope of the $H$-distribution function~(Fig.~\ref{f:figlHm}).
Asteroids smaller than about 30--50 km in diameter are thought to have evolved significantly over the age of the solar system, both in their mass and in their orbital parameters, due to high velocity collisional events and to non-gravitational solar
radiation effects which cause stochastic orbital drift; larger asteroids are thought to be little perturbed in their orbital parameters since the time of the last major dynamical events in the early history of the solar system~\citep{Bottke:2015}.
Therefore, we define the following two subsets of large asteroids:
\begin{enumerate}
\item The ``nearly primordial asteroids'', defined as those brighter than absolute magnitude $H=10.8$ (equivalently, diameter $D\gtrsim30$~km); these asteroids are indicated in Fig.~\ref{f:figpae} with the black dots, and their sample size is 925.
(We note that this set of MBAs is only $\sim14\%$ larger than the set studied by Plummer, indicating that this set of bright asteroids was observationally nearly complete a century ago.)

\item The ``even more primordial asteroids'', brighter than absolute magnitude $H=8.2$ (equivalently, diameter $D\gtrsim100$~km).  This is a smaller sample of 158 asteroids; they are indicated by the blue circles in Fig.~\ref{f:figpae}.
\end{enumerate}
The equivalent diameters quoted above are based on an assumed mean visual albedo, $0.09$. While the absolute magnitudes $H$ are fairly securely determined for nearly all asteroids, only a subset have well-determined albedos; moreover, the albedos vary with taxonomic-type and have large scatter~\citep{Masiero:2011,Usui:2013}.  The adopted albedo value is an approximate mean value for the main belt asteroids.

It is also interesting to examine the eccentricity distribution as a function of heliocentric distance.  Many previous studies of the asteroid belt define an inner, middle and outer belt with the following boundaries in semimajor axis: $(2.06<a/\AU<2.5), (2.5<a/\AU\leq2.82)$ and $(2.82<a/\AU\leq3.27)$, respectively.  With these definitions, the inner boundary of the inner belt is near the 4:1 mean motion resonance (MMR) with Jupiter (and it excludes the innermost group of asteroids known as the Hungarias), the boundary between the inner and middle belt is at the 3:1 MMR with Jupiter, that between the middle and outer belt is at the 5:2 MMR with Jupiter, and the outer boundary of the outer belt is at the 2:1 MMR with Jupiter.  For the observationally complete MBAs $(H\leq15)$, the sample sizes of these three regions are 13263, 21325, and 28702, respectively. 

\section{Analysis}\label{s:analysis}

\subsection{Testing for a  Rayleigh distribution}\label{s:testingRayl}

\begin{figure}
\centering\includegraphics[scale=0.33,angle=270]{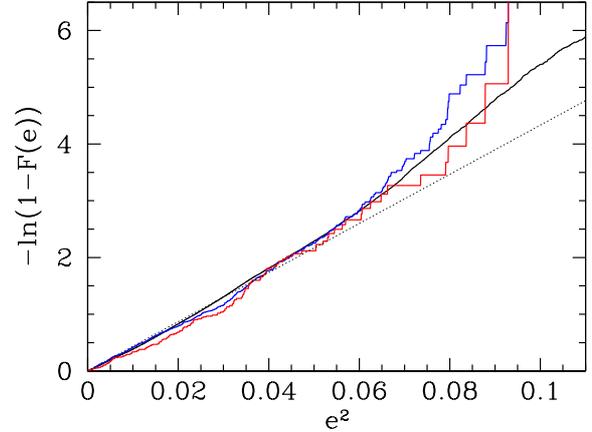}
\caption{
Comparison of the eccentricity distribution as a function of asteroid size. The black curve is for the observationally complete set of MBAs ($H\leq15$), the blue and red curves are for the primordial set ($H\leq10.8$) and the even more primordial set ($H\leq8.2)$, respectively. 
Plot of $-\ln(1-F(e))$ as a function of $e^2$, where $F(e)$ is the cumulative fraction of asteroids with eccentricity $<e$.
 The dotted straight line represents a Rayleigh distribution with parameter $\sigma=0.106$.
}
\label{f:figR}
\end{figure}

To investigate whether the MBAs' eccentricities follow the Rayleigh distribution, it is useful to note that the cumulative distribution function (CDF) of the Rayleigh distribution is $F(e)=1-\exp(-e^2/(2\sigma^2))$.  Thus, if the eccentricities obeyed the Rayleigh distribution, a plot of $-\ln(1-F(e))$ versus $e^2$ would be a straight line with slope $(2\sigma^2)^{-1}$.  Fig.~\ref{f:figR} provides this plot for the observationally complete MBAs (in black).
We observe that the eccentricities deviate from a straight line.   
Also shown in Fig.~\ref{f:figR} are the plots for the large asteroids, in blue for the nearly primordial set ($H\leq10.8$) and in red for the even more primordial set ($H\leq8.2)$.  We observe that the deviation from a Rayleigh distribution persists for the larger asteroids, although the statistics is poorer.  We performed the two-sample Kolmogorov-Smirnov test and found p-values exceeding 0.05 for each pair of samples, $(H\leq15,H\leq10.8)$ and $(H\leq15,H\leq8.2)$ and $(H\leq10.8,H\leq8.2)$, indicating that the samples are consistent with being drawn from the same distribution.

\begin{figure}
\centering\includegraphics[scale=0.45]{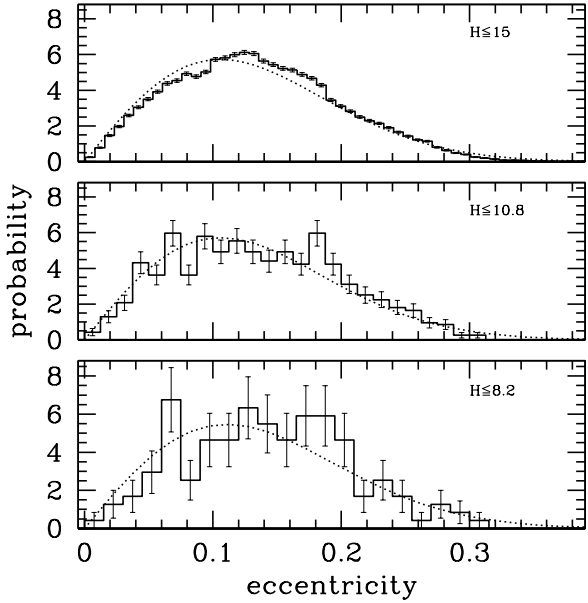}
\caption{
Histograms of the eccentricities for three size dependent samples of MBAs: the observationally complete set of MBAs ($H\leq15, D\gtrsim5$~km) is shown in the top panel, the nearly primordial set ($H\leq10.8, D\gtrsim30$~km) is in the middle panel, and the even more primordial set ($H\leq8.2, D\gtrsim100$~km) is in the bottom panel. The error bars indicate Poisson uncertainties. In each case, the dotted curves plot the best-fit Rayleigh distribution. 
}
\label{f:fig3se}
\end{figure}

We plot in Fig.~\ref{f:fig3se} the histograms of the MBAs' eccentricities for the three size dependent samples.  The mean eccentricities are 0.135, 0.135 and 0.142 for the samples with $H\leq15, ~H\leq10.8$ and $H\leq8.2$, respectively, and their standard deviations are 0.065, 0.065 and 0.064, respectively. Also plotted are the best-fit Rayleigh distributions in each case; the best-fit Rayleigh parameter is found to be $0.106, 0.106, 0.111$, respectively.  Qualitatively, the deviation from the best-fit Rayleigh distribution can be described as a deficit of small eccentricities, $e\lesssim0.1$, and excess of higher eccentricities.


\begin{figure}
\centering\includegraphics[scale=0.33,angle=270]{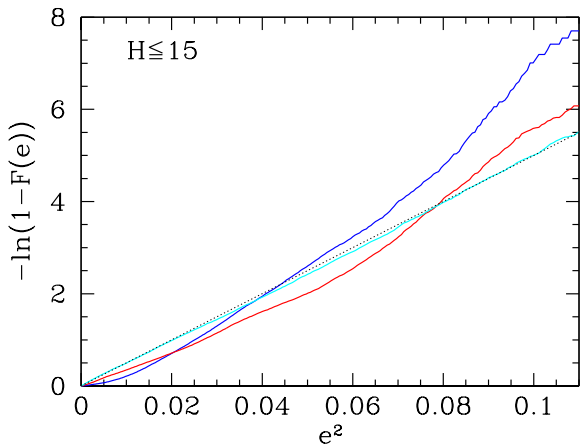}
\caption{
Comparison of the eccentricity distributions in the inner $(2.06<a/\hbox{AU}\leq2.5$, in blue), middle $(2.5<a/\hbox{AU}\leq2.82$, in red) and outer belt $(2.82<a/\hbox{AU}\leq3.27$, in cyan).  Plot of $-\ln(1-F(e))$ as a function of $e^2$, where $F(e)$ is the cumulative fraction of asteroids with eccentricity $<e$.
The dotted straight line represents a Rayleigh distribution with parameter $\sigma=0.101$.
}
\label{f:figsyn3}
\end{figure}

\begin{figure}
\centering\includegraphics[scale=0.45]{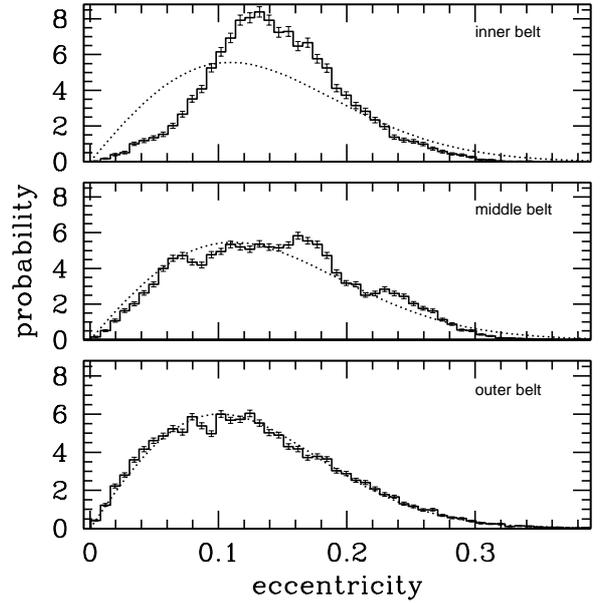}
\caption{
Histograms of the eccentricities in the inner $(2.06<a/\hbox{AU}\leq2.5$, top panel), middle $(2.5<a/\hbox{AU}\leq2.82$, center panel) and outer belt $(2.82<a/\hbox{AU}\leq3.27$, bottom panel). The error bars indicate Poisson uncertainties. The dotted curves plot the best-fit Rayleigh distributions, which have parameter $\sigma=0.109,0.111,0.101$ for the inner, middle and outer belt, respectively. 
}
\label{f:fig3e}
\end{figure}

We carry out a similar analysis for the heliocentric distance dependence of the eccentricities.
The plots of $-\ln(1-F(e))$ versus $e^2$ for the inner, middle and outer asteroid belt are shown in Fig.~\ref{f:figsyn3}, while histograms of the eccentricities are shown in Fig.~\ref{f:fig3e}.  The mean eccentricities are 0.145, 0.143 and 0.126, for the inner, middle and outer asteroid belt, respectively; the standard deviations are 0.052, 0.067 and 0.067, respectively.  Although the mean and standard deviation values are not significantly different, by visual inspection we find that there are significant differences in the shapes of the eccentricity distributions of these samples.  The inner and middle belts deviate significantly from the Rayleigh distribution; in particular, the inner belt has a large deficit at small eccentricities, $e\lesssim0.1$, and excess at higher eccentricities, relative to the best-fit Rayleigh distribution.  However, the outer belt follows the Rayleigh distribution fairly closely, with parameter $\sigma=0.101$.  These observations are confirmed with Kolmogorov-Smirnov two-sample tests which show that the three samples differ from each other.

\subsection{Comparison with Plummer (1916)}\label{s:compPlummer}

At first sight, our result for the large asteroids appears to differ from that of \cite{Plummer:1916} obtained a century ago, when the sample of large asteroids, $H\leq10.8$, was already $\sim85\%$ complete.   What could be the cause of the difference?  
A possible cause is that Plummer considered osculating elements whereas we have considered proper elements.  

\begin{figure}
\includegraphics[scale=0.33,angle=270]{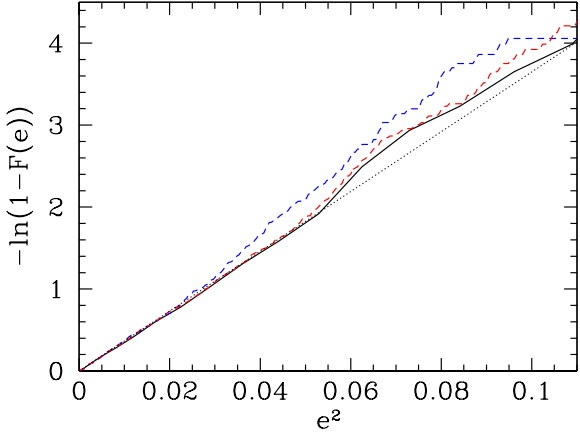}
\caption{Plots of $-\ln(1-F(e))$ as a function of $e^2$, where $F(e)$ is the cumulative fraction of asteroids with eccentricity $<e$.
The blue and red dashed-line curves are for the proper and osculating eccentricities, respectively, of the 809 lowest-numbered MBAs, the black curve is for the data used by Plummer (1916).  The dotted straight line represents Plummer's estimated Rayleigh distribution with parameter $\sigma=0.117$.
}
\label{f:figPlummer}
\end{figure}

Although we do not have the exact data that Plummer worked with, he did present the binned data of the 809 eccentricities (taken from \cite{Utzinger:1915}); we plot the distribution, $-\ln(1-F(e))$, for this data in Fig.~\ref{f:figPlummer}. 
For comparison with current data we can approximate the century-old observational sample by examining the current data of the 809 lowest-numbered asteroids (because the numbering convention is strongly correlated with discovery epoch for the early discoveries of minor planets).  The distributions of the osculating and proper eccentricities of these 809 asteroids are also plotted in Fig.~\ref{f:figPlummer}.  
We find that our conjecture does indeed provide a plausible explanation:  the distribution of the osculating eccentricities of this sample shows smaller deviation from the Rayleigh distribution than the proper eccentricities; moreover, the small deviations are similar to those of Plummer's sample.  (The small differences between Plummer's data and the current data of the osculating eccentricities of the 809 lowest-numbered MBAs are possibly owed to improvements in the orbital parameters of these MBAs, and, to a lesser extent, a small number of individual MBAs that are not common to the two samples.) 
Plummer dismissed these deviations as being ``of a kind which could scarcely be avoided by any practical frequency function''; in other words, he attributed the deviations to statistical fluctuations.  This is a reasonable conclusion for the osculating eccentricities of this relatively small sample.  However, the deviations from the Rayleigh distribution are obviously larger for the proper eccentricities.  A Kolmogorov-Smirnov test confirms that the proper eccentricities are not consistent with a Rayleigh distribution.

The deviations from Plummer's eccentricity distribution function were already noted by \cite{Beck:1981} when the catalog of minor planets had increased in number to 2176.  Beck's analysis was also based on the osculating elements, the proper elements not yet having gained attention at that time.  He concluded that the eccentricities appeared to be independent of size (over the one-order-of-magnitude range in diameter spanned by the data). He also noted hints in the data of an excess of high eccentricities, $e\gtrsim0.3$ (which he attributed to inclusion of comets in the sample and/or observational selection bias), as well as hints of a double maximum, each peak possibly related to heliocentric distance.  He considered the size of the sample too small to conclude that these features were statistically significant.

\subsection{Model distributions}\label{s:models}

\begin{table*}
\begin{minipage}{126mm}
\caption{Best-fit model parameters}
\begin{tabular}{|l||c|c|c|c|}
\hline
 & \multicolumn{2}{|c|}{Model 1} &\multicolumn{2}{|c|}{Model 2}\\
  \cline{2-5}
MBA sample & $\bar\sigma$ &  $\bar\delta_e$ & $\bar{e_c}$ & $\bar\delta_e$  \\
  \hline  
  $H\le15$     & $0.0786\pm0.0009$ & $0.1010\pm0.0015$ & $0.0338\pm0.0009$  & $0.113\pm0.005$  \\
  $H\le10.8$  & $0.0784\pm0.0041$   & $0.101\pm0.007$     & $0.0301\pm0.002$    & $0.114\pm0.006$   \\
  $H\le8.2$    & $0.0728\pm0.0097$ & $0.115\pm0.016$     & $0.0251\pm0.005$    & $0.133\pm0.014$  \\
   \hline
 inner belt & $0.0559\pm0.0023$ & $0.133\pm0.004$ & $0.0239\pm0.0014$ & $0.134\pm0.007$  \\
 middle belt  & $0.0772\pm0.0025$  & $0.115\pm0.005$ & $0.0358\pm0.0020$ & $0.121\pm0.008$   \\
 outer belt  & $0.0974\pm0.0047$ & $0.030\pm0.026$ & $0.0365\pm0.0014$ & $0.095\pm0.005$ \\
\hline 
\end{tabular}
\label{Tbl1}
\end{minipage}
\end{table*}

Asteroids are expected to have formed on nearly circular co-planar orbits~\citep[see, e.g., review by][]{Johansen:2015}.  However, the current eccentricities in the main asteroid belt have long been recognized to be far in excess of those expected during their formation, and likely owed to post-formation dynamical excitation mechanisms such as secular resonance sweeping during the dissipation of the gaseous solar nebula or gravitational scattering action of long-lived planetary embryos or planetesimal-driven giant planet migration~\citep[see reviews by][]{Petit:2002,Morbidelli:2015}.  A satisfactory quantitative model of the origin of the asteroids' dynamical excitation remains elusive at this time.
Here, we carry out a simple model-fitting exercise: we assume that the current eccentricity of an asteroid is the vector sum of an initial eccentricity and an excitation, as follows:
\begin{equation}
	{\bf e}_f = {\bf e}_i+\vdelta_e,
\label{e:ef}\end{equation}
where ${\bf e}=e(\cos\varpi,\sin\varpi)$ is the usual eccentricity vector with $\varpi$ being the longitude of perihelion, and the subscripts $i$ and $f$ indicate values long before and long after the dynamical excitation. We identify the current proper eccentricities with $e_f$.  We assume that the initial eccentricities, $e_i$, prior to the excitation, followed a theoretically expected distribution function (the specific functions are discussed below). We also assume that the proper longitudes of perihelion, both initial and final, have a uniform random distribution.  The magnitude of the excitation term, $\delta_e$, is assumed to be a free parameter, and its phase, is assumed randomly distributed.  This model is motivated by analytical theories of the effects of secular resonance sweeping either during the dissipation of the solar nebula or during the planetesimal-driven orbital migration of the giant planets~\citep{Ward:1976,Gomes:1997,Minton:2011}. 

For the initial eccentricities, we consider two distribution functions.  
The first is a Rayleigh distribution, predicted by theoretical and numerical studies of self-gravitating planetesimal belts~\citep[e.g.,][]{Ida:1992,Greenzweig:1992,Ohtsuki:2000},
\begin{equation}
p_R(e) = {e\over\sigma^2}\exp(-{e^2\over2\sigma^2}),\label{e:Ray}
\end{equation}
where the parameter $\sigma$ is the mode of the distribution.
Alternatively, in the Kepler-shear dominated regime, the heavier-tailed 
two-dimensional Cauchy distribution is expected~\citep{Collins:2007},
\begin{equation}
p_C(e) = {e_ce\over (e^2+e_c^2)^{3/2}},\label{e:nonRay}
\end{equation}
where $e_c$ is the mode of the distribution.

If the initial eccentricities, $e_i$, follow a Rayleigh distribution as in Eq.~(\ref{e:Ray}), then the probability distribution function and cumulative distribution function of $x=e_f\cos\varpi_f$ (or $e_f\sin\varpi_f$) can be expressed as
\begin{align}
    f_R(x,\sigma,\delta_e)={1\over \sigma\sqrt{2\pi^3}} \int_{-1}^{1}{\rm{d}}u\, (1-u^2)^{-{1\over2}}\exp{{-(x-u\delta_e)^2}\over {2\sigma^2}},\label{e:pdfRay}\\
    F_R(x,\sigma,\delta_e)={1\over 2} + {1\over 2\pi}\int_{-1}^{1}{\rm{d}}u\, (1-u^2)^{-{1\over2}}{\rm{erf}}({{x-u\delta_e}\over \sqrt{2}\sigma}).\label{e:cdfRay}
\end{align}
We treat $\sigma$ and $\delta_e$ as free parameters.
Using the data of the proper eccentricities, $e_f$, we generate random numbers for their proper longitudes of perihelion, $\varpi_f$, and hence generate the data sample for $x$.  We fit this data to the model (Eq.~\ref{e:pdfRay}-\ref{e:cdfRay}) and find the best-fit values of $\sigma$ and $\delta_e$. This procedure is repeated $100$ times to generate 100 random realizations of the data sample and, in each case, we find the best-fit values of $\sigma$ and $\delta_e$.  We then compute the average values  $\bar\sigma$ and $\bar\delta_e$.

If the initial eccentricities, $e_i$, follow the two dimensional Cauchy distribution as in Eq.~(\ref{e:nonRay}), then the probability distribution function and cumulative distribution function of $x=e_f\cos\varpi_f$ (or $e_f\sin\varpi_f$) can be derived as
\begin{align}
    f_{C}(x,e_c,\delta_e)={1\over\pi^2}\int_{-1}^{1}{\rm{d}}u\, (1-u^2)^{-{1\over2}} {e_c \over {{e_c}^2 + (x-u\delta_e)^2} },\label{e:pdfnonRay}\\
    F_{C}(x,e_c,\delta_e)={1\over 2} + {1\over\pi^2}\int_{-1}^{1}{\rm{d}}u\, (1-u^2)^{-{1\over2}}\arctan({{x-u\delta_e}\over e_c}).\label{e:cdfnonRay}
\end{align}
We treat $e_c$ and $\delta_e$ as free parameters, and we find their best-fit values for this model.  Using the method described above, we obtain the average values $\bar e_c$ and $\bar\delta_e$. 

Below we refer to the model with the Rayleigh distribution of initial eccentricities (described by Eq.~\ref{e:pdfRay}-\ref{e:cdfRay}) as ``Model 1", and we refer to the model with the two dimensional Cauchy distribution of initial eccentricities (described by Eq.~\ref{e:pdfnonRay}-\ref{e:cdfnonRay}) as ``Model 2".  
We carried out these model-fits for the MBAs with the three different cut-offs in absolute magnitude ($H\le15,\,H\le10.8,\, H\le8.2$) and also for the inner, middle and outer belt MBAs.  The parameters of the best-fit models in each case are given in Table 1.

We mention that, because asteroid eccentricities are limited to $0\leq e\leq 1$, the distribution functions $p_R(e)$ and $p_C(e)$ (Eq.~\ref{e:Ray}--\ref{e:nonRay}) must be truncated to this domain.  Such truncation affects the normalization factors of $p_R(e)$ and $p_C(e)$, and also introduces complexity in the derivation of the probability density functions and cumulative density functions of $e_f(\cos\varpi_f,\sin\varpi_f)$ for each model.  Extending the upper limit of $e$ to infinity in Eq.~(\ref{e:Ray})--(\ref{e:nonRay}) greatly simplifies these derivations with only a small penalty in numerical accuracy.  When we carry out the model-fitting with the truncated distributions, the best-fit parameters lie well within one standard deviation of those listed in Table 1.

\subsection{Size dependence}\label{s:sizedependence}

\begin{figure}
\centering
\hglue-0.2truein\includegraphics[width=250px]{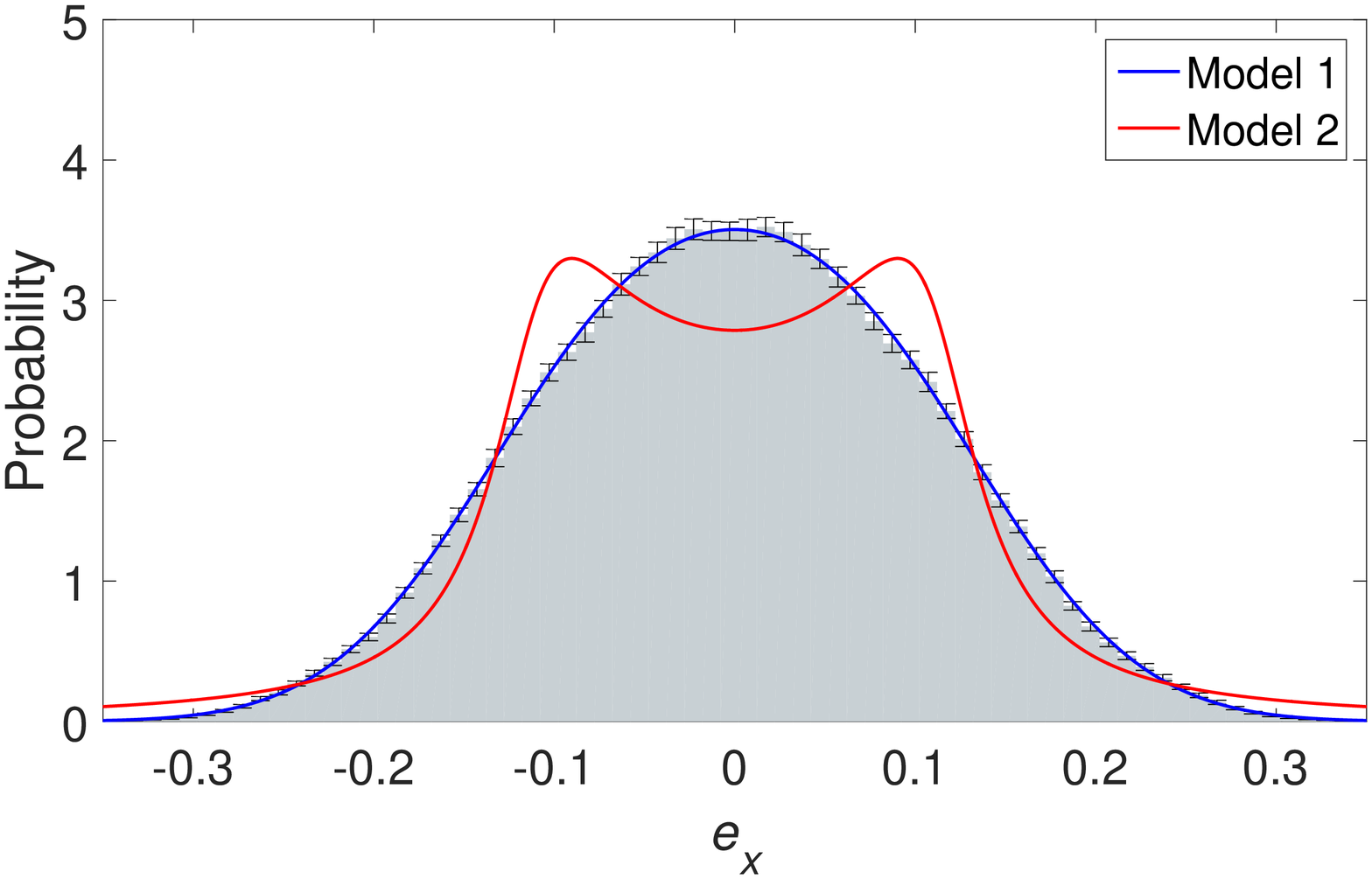}
\hglue-0.2truein\includegraphics[width=250px]{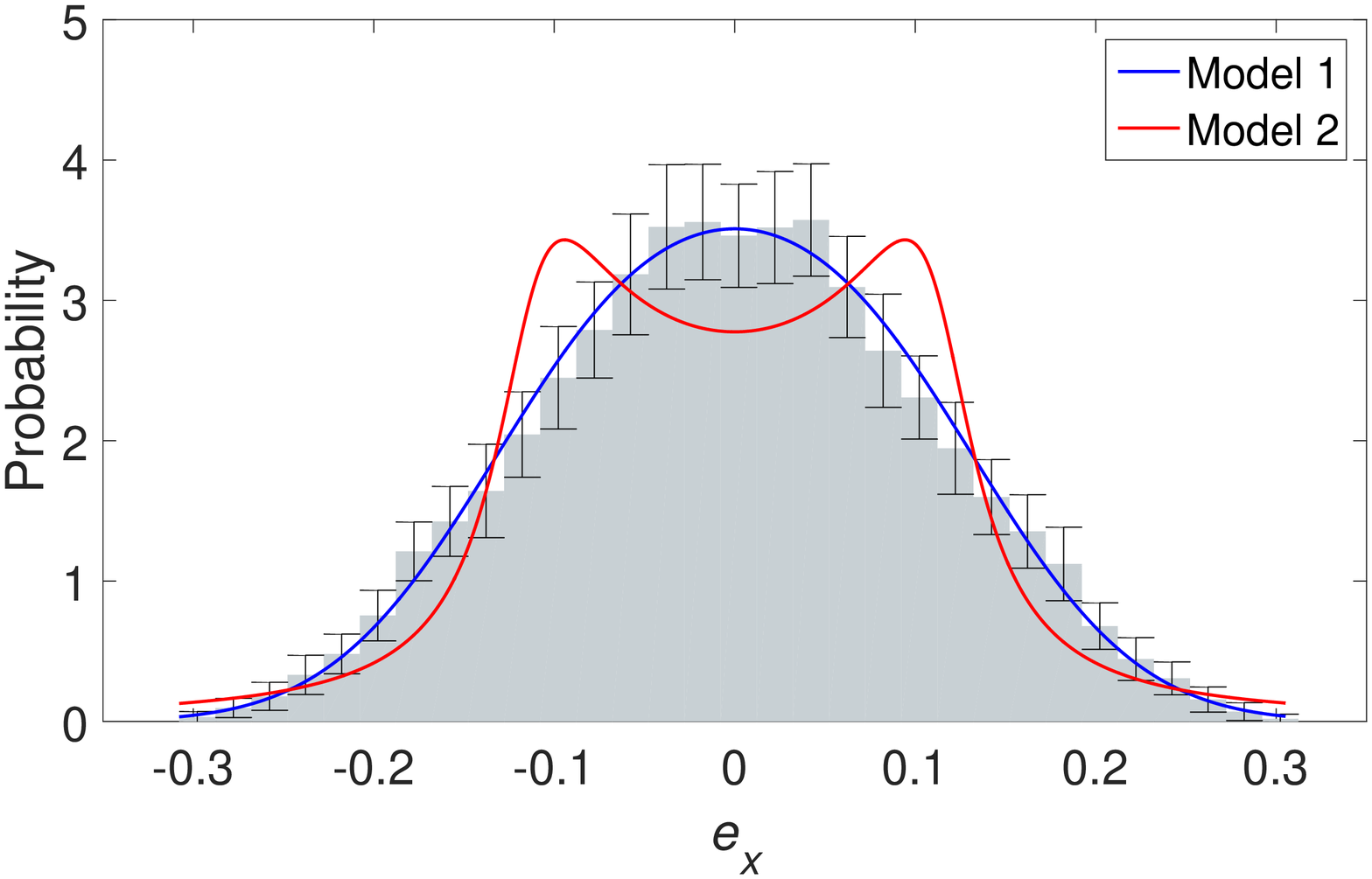}
\hglue-0.2truein\includegraphics[width=250px]{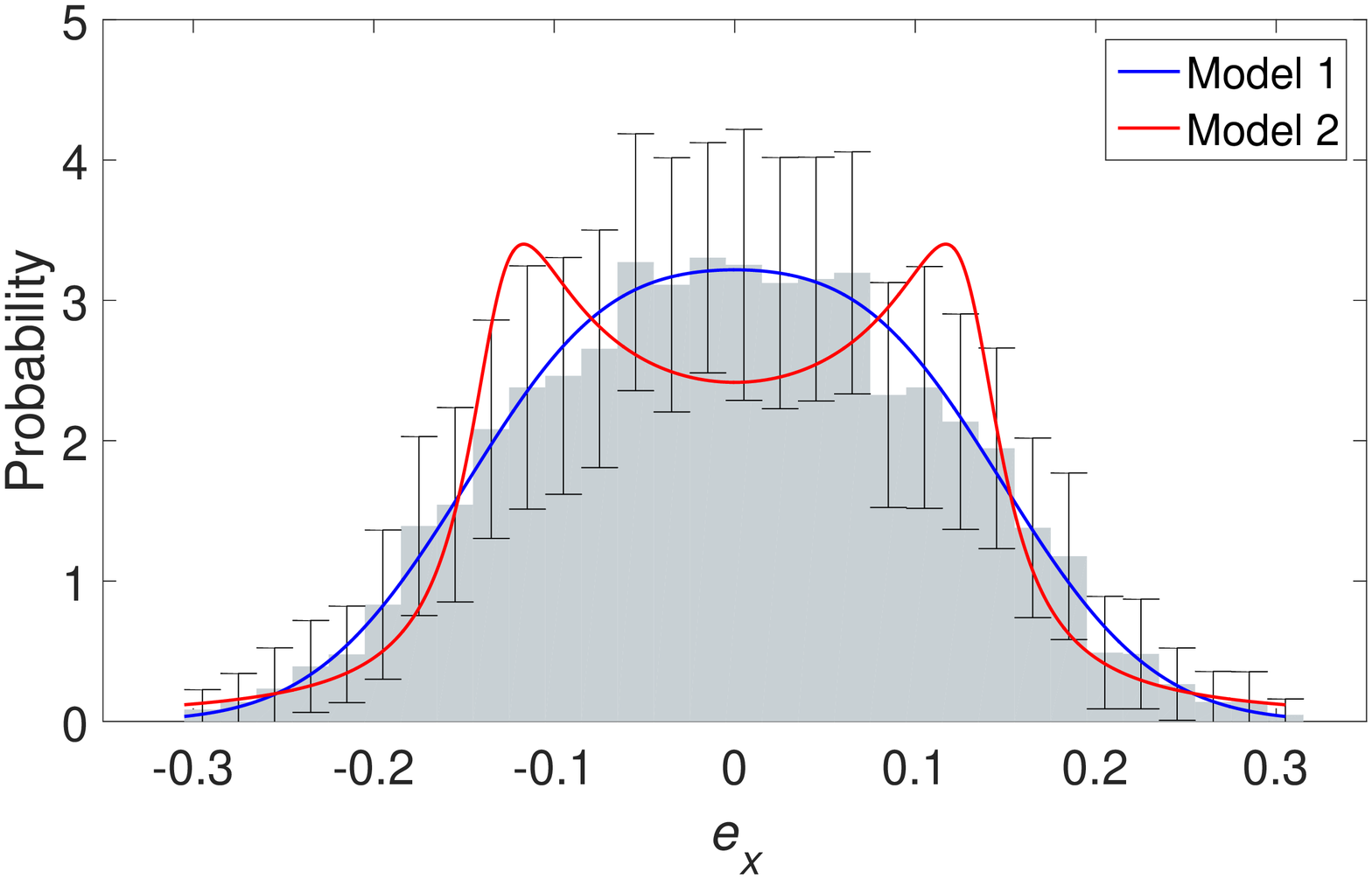}
\hglue+0.3truein\caption{Distribution of the proper eccentricity vector components, $e_x=e_f\cos\varpi_f$ for the observationally complete set of MBAs with $H\le15$ (top), the nearly primordial set with $H\le10.8$ (center), and the ``even more primordial'' set of $H\le8.2$ (bottom). The shaded histogram is the average of the $100$ realizations of the random values of $\varpi_f$; the error bars indicate their standard deviation in each bin.  The blue and red curves are for Model 1 and Model 2, respectively, with parameter values listed in Table 1. 
}\label{f:threeH}
\end{figure}

Examining the results in Table 1 for the three size dependent samples, ($H\le15,\,H\le10.8,\,H\le8.2$), we find that the best-fit model parameters of the larger MBAs do not differ dramatically from those of the observationally complete set of MBAs.  
We also find that the typical pre-excitation eccentricities revealed by Model 1 are noticeably larger than those revealed by Model 2 ($\sim0.08$ versus $\sim0.03$), whereas the magnitude of the excitation parameter, $\delta_e$, is similar in both models ($\sim 0.1$). 

For the three size dependent samples, we plot in Fig.~\ref{f:threeH} the distributions of $e_x=e_f\cos\varpi_f$.  (The distributions of $e_y=e_f\sin\varpi_f$ are statistically identical.) The shaded histogram is the average of 100 realizations of the random values of $\varpi_f$; the error bars indicate the standard deviation of the 100 realizations.  (The standard deviations are larger for the larger asteroids, reflecting the decreasing sample sizes hence larger statistical fluctuations.)  Overall, we observe that the histograms of the three samples have fairly similar shapes. This is consistent with the finding above that the best-fit model parameters are not very sensitive to the asteroid sizes.

Also plotted in Fig.~\ref{f:threeH} are the best-fit Model 1 (in blue) and the best-fit Model 2 (in red) for each data sample.  Visual inspection finds that Model 1 fits the averaged histograms better than Model 2.  We performed the Anderson-Darling test to evaluate the models' goodness of fit for each of the 100 realizations of the random values of $\varpi_f$.  For the asteroids of $H\leq15$, we find that most of the p-values are smaller than 0.05 for both Model 1 and Model 2, indicating that these models are not a good fit.  For the asteroids of $H\leq10.8$, we find that most p-values for Model 1 are greater than 0.05 but most p-values for Model 2 are less than 0.05.  For the sample of the largest asteroids, $H\leq8.2$, we find that most of the p-values for both models exceed 0.05, indicating that we cannot reject either model for this sample.   

We interpret these statistical goodness-of-fit results as follows.  The small sample size of the largest asteroids, $H\leq8.2$, does not allow us to reject either model, and both models appear to be reasonable fits to the data.  Recalling that all three samples' eccentricity distributions are consistent with being drawn from the same intrinsic distribution (see Section~\ref{s:testingRayl}), the larger samples should provide a more discriminating test of the models.  The larger sample size of $H\leq10.8$ allows us to reject Model 2 but not Model 1.  The vastly larger sample size of $H\leq15$ is large enough that, even though visual inspection indicates that Model 1 is a good fit to the data, the statistical test of goodness of fit rejects this model because the visually small deviations of the data from the model are nevertheless statistically significant.  We conclude that while Model 1 provides a good description of the data, the data contain more information than can be captured by this simple model. 

\subsection{Heliocentric distance dependence}\label{s:distancedependence}

\begin{figure}
\centering
\hglue-0.2truein\includegraphics[width=250px]{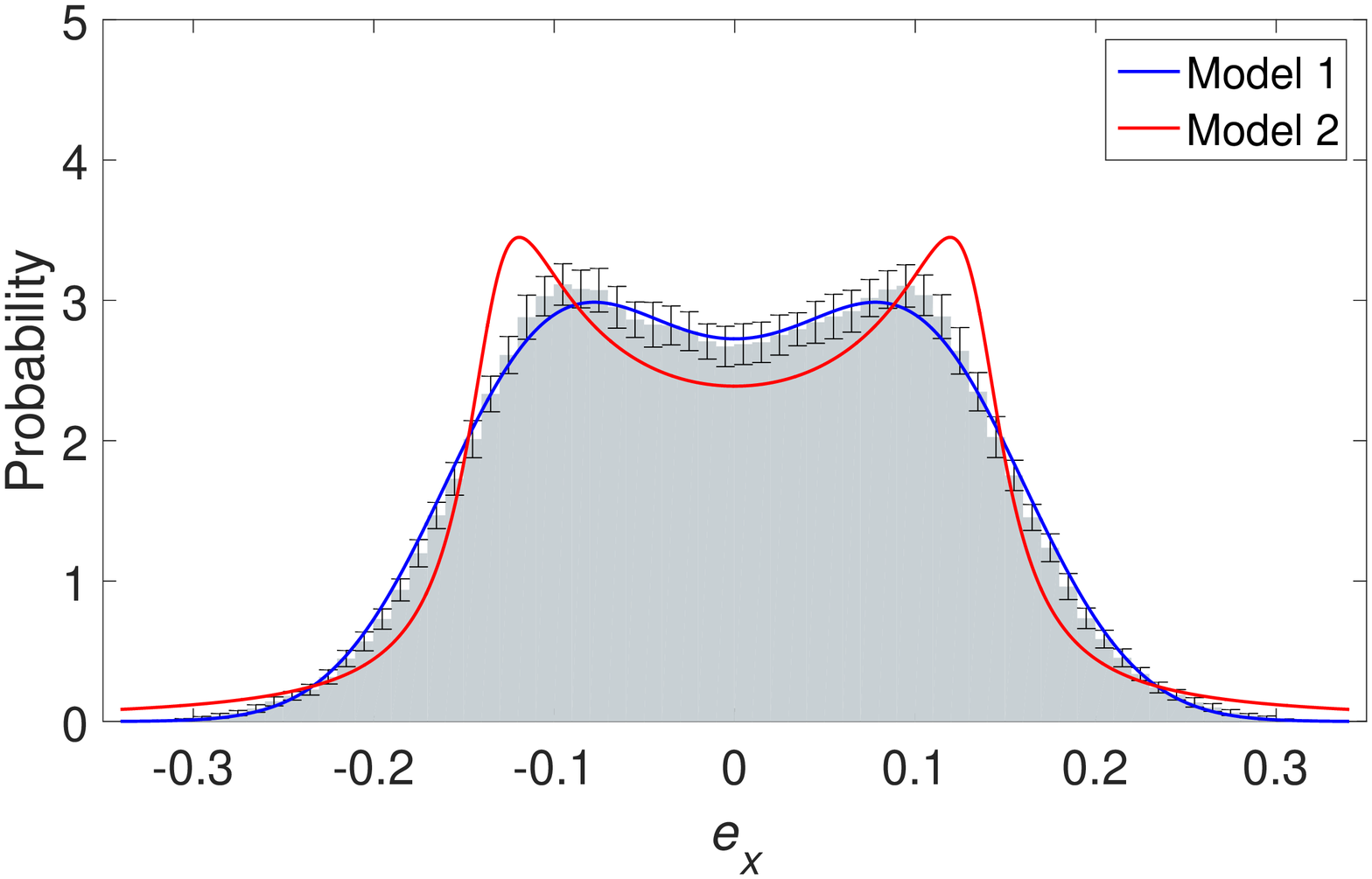}
\hglue-0.2truein\includegraphics[width=250px]{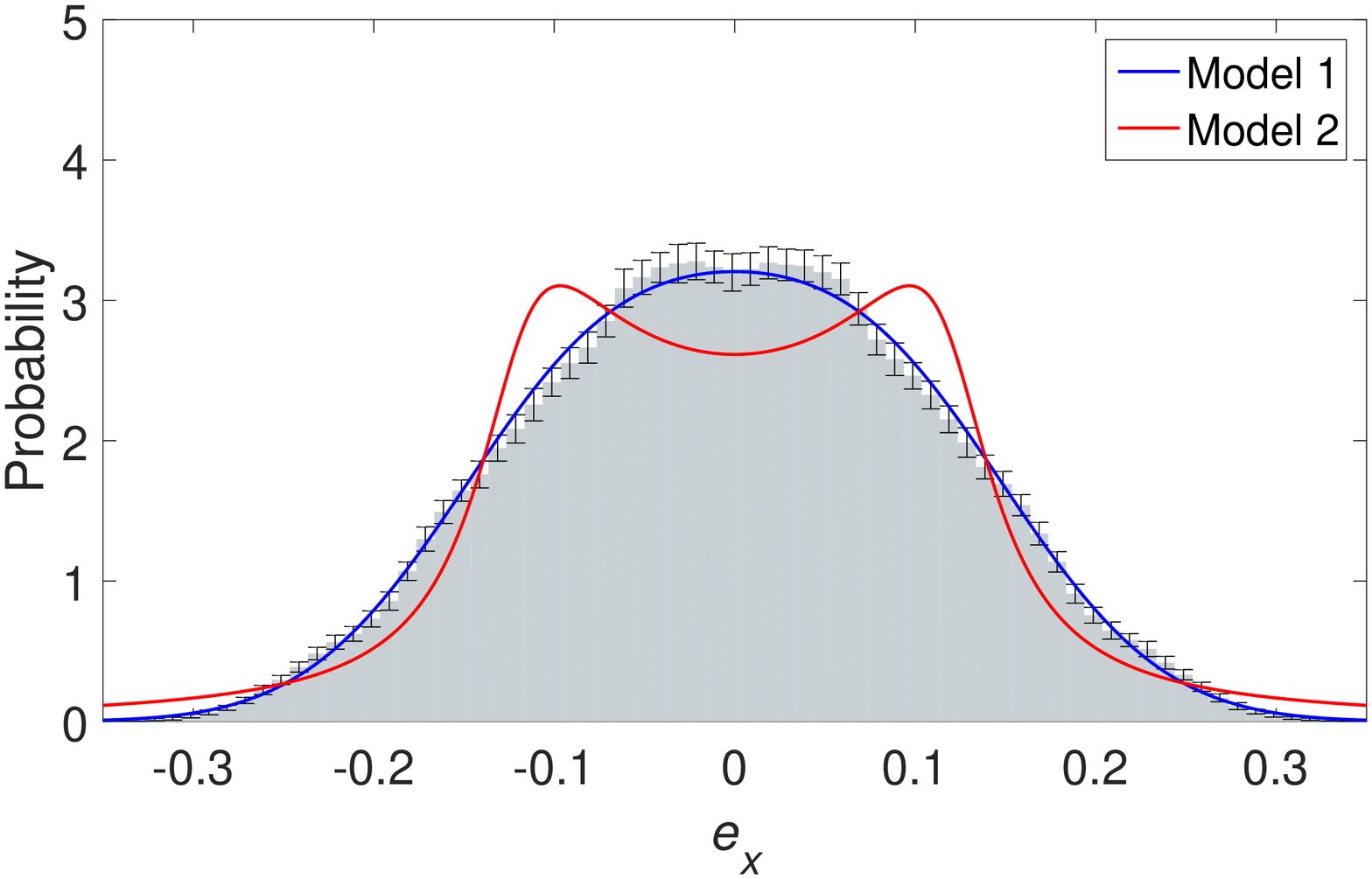}
\hglue-0.2truein\includegraphics[width=250px]{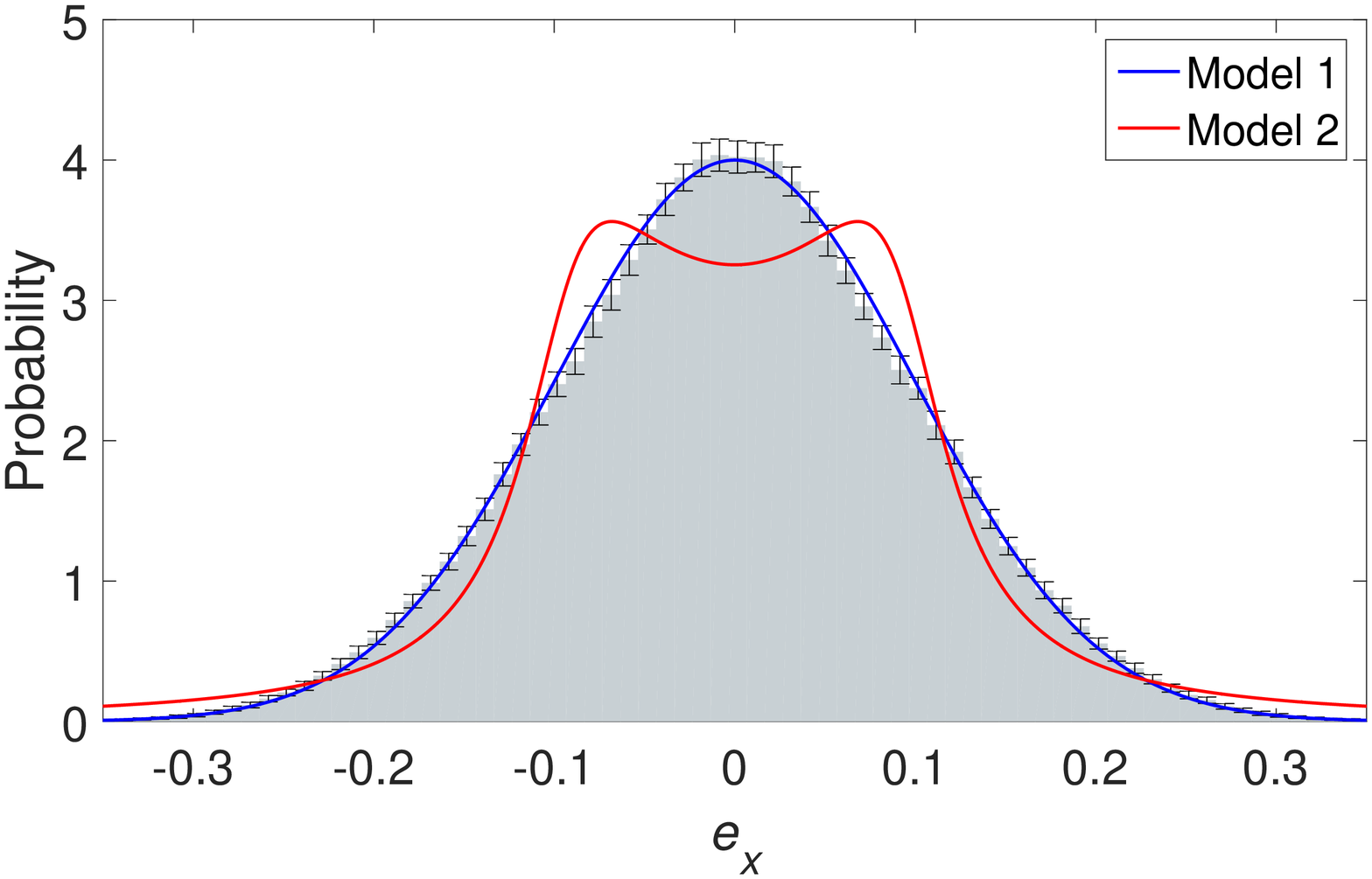}
\hglue+0.3truein\caption{Distribution of the proper eccentricity vector components, $e_x=e_f\cos\varpi_f$ for the inner belt (top), the middle belt (center), and the outerbelt (bottom). The shaded histogram is the average of the $100$ realizations of the random values of $\varpi_f$; the error bars indicate their standard deviation in each bin.  The blue and red curves are for Model 1 and Model 2, respectively, with parameter values listed in Table 1. 
}\label{f:threebelts}
\end{figure}

Examining the results in Table 1 for the dependence on heliocentric distance, we find that for both Model 1 and Model 2 the inner belt has significantly smaller values of the pre-excitation eccentricities and larger values of the excitation parameter $\delta_e$ compared with the outer belt; the middle belt has parameter values intermediate between the inner and outer belt parameters.   

We plot in Fig.~\ref{f:threebelts} the distributions of $e_x=e\cos\varpi$ for the three heliocentric distance dependent sets of MBAs (inner, middle and outer belt). As before, the shaded histogram is the average of 100 realizations of the random values of $\varpi$, and the standard deviation of the 100 samples is indicated by the error bars.  We observe striking differences in the distributions of $e_x$ as a function of heliocentric distance. The inner belt sample is obviously double peaked, with peaks of $e_x$ near $\pm0.1$; there is also a hint of secondary peaks near $\pm0.03$.  The middle belt has a single peak but the peak is somewhat flat, hinting of an underlying double peaked distribution with insufficient separation of the two peaks.  The outer belt has a single peaked distribution. The latter is unsurprising when we recall our finding in Section~\ref{s:testingRayl} that the outer belt eccentricities follow quite well a Rayleigh distribution, their eccentricity vector components are then expected to follow well a Gaussian distribution with zero mean.  

Also plotted in Fig.~\ref{f:threebelts} are the best-fit Model 1 and best-fit Model 2.  By visual inspection we find that, for all three samples, Model 1 fits the averaged histograms much better than Model 2. However, there are small but noticeable deviations of the data from Model 1.  We performed the Anderson-Darling test to evaluate the models' goodness of fit.  In all cases, we find p-values less than 0.05, indicating that we should reject both models for all three samples (the inner, middle and outer belts).  However, it is evident that these low p-values signal only the statistical significance of the small deviations from Model 1, whereas the most prominent feature, the double-peaked distribution of $e_x$ (and $e_y$) in the inner belt (and to a lesser extent the middle belt), is described quite well by this model. The data sample sizes are large enough that more sophisticated models should be investigated; we leave that to future work.

\section{Summary and Discussion}\label{s:summary}

The observationally complete sample size of the main belt asteroids (excluding members of collisional families) is now more than 64,000, a nearly 80-fold increase over the past century.  We undertook a careful statistical analysis of the eccentricities of this sample and their interpretation with simple physical models. We obtained the following results.  

\begin{enumerate}

\item Plummer's (1916) conclusion that the eccentricities of the then known 809 main belt asteroids obeyed a Rayleigh distribution holds for the {\it osculating} eccentricities of large asteroids, but not for their proper eccentricities.

\item The proper eccentricity distribution does not depend significantly on asteroid size over the range spanned by the observationally complete MBAs, i.e., diameter 5~km~$\lesssim D\lesssim950$~km, more than two orders of magnitude in asteroid size. 

Asteroids of absolute magnitude $H\leq8.2$ (equivalent diameter $D\gtrsim100$~km), those of absolute magnitude $H\leq10.8$ ($D\gtrsim30$~km), and those of absolute magnitude $H\leq15$ ($D\gtrsim5$~km), are all consistent with being drawn from the same intrinsic distribution of proper eccentricities.
All of these samples deviate from a Rayleigh distribution.  There is a deficit of eccentricities smaller than $\sim0.1$ and an excess of larger eccentricities. 

\item 
The proper eccentricity distribution depends significantly on heliocentric distance.
The inner, middle and outer belts, defined with boundaries in proper semimajor axis $(2.06<a/\AU<2.5), ~(2.5<a/\AU\leq2.82)$ and $(2.82<a/\AU\leq3.27)$, respectively, differ from each other. The outer belt follows a Rayleigh distribution, but the inner belt and middle belt do not. 
 
 \item The inner belt has a striking double peaked distribution of the proper eccentricity vector components, $e(\cos\varpi,\sin\varpi$). The outer belt does not show evidence of such a feature, and the middle belt shows a hint of such a feature.

 \item The proper eccentricity vectors in the inner belt can be modeled quite well as a vector sum of an initial eccentricity of random phase and magnitude drawn from a Rayleigh distribution of parameter $\sim0.06$, and an excitation of random phase and magnitude $\sim0.13$.  However, the sample size is large enough that more sophisticated model-fitting is warranted in a future investigation.

\end{enumerate}

It is not inconceivable that unrecognized collisional family members contribute to the non-trivial features of the eccentricity distributions that we have found.  We adopted the family identifications as reported in the AstDys-2 database.  Our results would need to be re-examined if the criteria for the identification of asteroid families were to be significantly revised. 

Our statistical analysis has revealed two trends in the MBAs' eccentricities which speak to their dynamical history and which have hitherto escaped attention.  The first is that the MBAs' eccentricities do not show any significant size dependence over more than two orders of magnitude in size, $\sim$~5--950~km.  This indicates that the mechanisms responsible for the dynamical excitation of the asteroid belt were size-independent. This also implies that, subsequent to their dynamical excitation, the smaller asteroids' eccentricities have evolved little, if at all, relative to the eccentricities of the larger asteroids.  Secondly, the MBAs' proper eccentricities show strong dependence on heliocentric distance, indicating that the dynamical excitation mechanisms were sensitive to heliocentric distance.  It is noteworthy that the mean eccentricity does not show a dramatic trend with heliocentric distance but the {\it shape} of the distribution function does. Previous studies using only mean or median eccentricity to characterize asteroids' eccentricities (and using only the small sample of large asteroids, $H\lesssim11$) overlooked this trend in the eccentricity distribution function.
 
Our model-fitting analysis, within the relatively simple models considered here (Eqs.~\ref{e:ef}--\ref{e:cdfnonRay}), leads to the following implications for the dynamical excitation history of asteroids.
Our finding that the shape of the eccentricity distribution function of the pre-excitation asteroid belt is described well with the Rayleigh distribution rather than the two-dimensional Cauchy distribution implies that (A) the pre-excitation asteroid belt resembled a gravitationally stirred disk of planetesimals, and (B) the timing of the excitation event(s) was not during the earliest epochs (when the disk would be in the Kepler-shear regime), but at a later epoch when the largest bodies had grown large enough that the population of bodies of the size of the observed asteroids, $D<1000$~km, was in the dispersion-dominated regime.  

In the dispersion-dominated pre-excitation asteroid belt, the eccentricities would be determined by a small number of the largest bodies, and are expected to be up to $\sim\!v_{\rm{esc}}/v_{\rm{orb}}$, where $v_{\rm{esc}}$ and $v_{\rm{orb}}$ is the surface escape velocity and the orbital velocity, respectively, of the largest bodies~\citep{Goldreich:2004}. 
In the inner belt, with an average orbital velocity of $\sim\!20$~km/s, the inferred pre-excitation eccentricities of $\sim\!0.056$ imply that largest bodies be of mass $m\approx0.8\times10^{25}$~g, diameter $D\approx1,700$~km (assuming a bulk density of $3,000$~kg~m$^{-3}$); in the outer belt, with a lower average orbital velocity of $\sim\!17$~km/s and higher inferred pre-excitation eccentricities of $\sim\!0.097$, the largest bodies would be a little larger, $m\approx2.6\times10^{25}$~g, $D\sim\!2,500$~km. 
With masses (1-3)$\times10^{25}$~g and sizes 1700--2500~kilometers, such bodies may be called ``planetary embryos''.  The required number of planetary embryos can be estimated as follows. These planetary embryos would have had nearly circular orbits (their eccentricities being damped by dynamical friction within the vast number of smaller bodies), and their gravitational influence would have extended to an annular region of width a few times their Hill radius, that is, $\sim kR_H= ka(m/3m_\odot)^{1/3}$, with $k\approx3$--10.  Thus their total number across the width, $\Delta a$, of the main asteroid belt would be $\sim\!\Delta a/kR_H$.  Assuming the same bulk density and the size estimates as above, we estimate approximately 35--100 planetary embryos would be necessary to account for the pre-excitation eccentricities of the MBAs. 
These estimates are smaller, both for the masses and the numbers of planetary embryos, than those found in modern numerical studies of the formation of the asteroid belt~\citep{Petit:2001,Chambers:2001b,Obrien:2007}.

Considering that the present-day asteroid belt contains not even one such body (the largest asteroid, Ceres, has diameter $D=946$~km, only about half as large as the planetary embryos' size estimates here), these bodies would need to be subsequently removed, possibly by their own mutual long-range gravitational perturbations or in combination with perturbations from the giant planets which would induce chaotic evolution and planetary collisions or ejections~\citep{Chambers:2001b}.  The implied numerical depletion exceeds a factor of (35--100).
Further, considering that the mass of the asteroid belt is dominated by the largest asteroids, the implied mass depletion by the excitation event(s) also exceeds a factor (35-100). This lower limit on the level of mass depletion is smaller than the factor (1000--6000) mass deficit of the asteroid belt inferred from other considerations, such as the smoothed radial density gradient estimates of the minimum mass solar nebula and models of the formation of asteroids~\citep{Weidenschilling:1977,Wetherill:1992,Chambers:2001b,Morbidelli:2009b}.  

It is of some importance to note that the relationship between the mass depletion factor and the numerical depletion factor is sensitive to the size distribution function at large sizes: to wit, the same mass depletion may be accomplished by a smaller numerical depletion if most of the mass is in a very few large bodies~\citep{Weidenschilling:2011}.  Thus, our estimate of a numerical depletion factor of $\gtrsim\!$~(35--100) combined with the independent estimates of a mass depletion factor of (1000--6000) derived in the above-mentioned previous studies may be useful in constraining the size distribution function of the large bodies, $D>1000$~km, in the primordial asteroid belt. We leave this to a future study.

Our model-fitting analysis also implies that the late excitation event(s) had little effect on the outer asteroid belt but had a large effect on the inner belt.  The double-peaked distribution of the proper eccentricity vector components in the inner belt is consistent with the effects of an inward sweeping of the $\nu_6$ secular resonance.  This secular resonance is now located near the inner edge of the asteroid belt but would have migrated inward at early times due to giant planet migration~\citep{Minton:2011}.  The absence of the double-peaked feature in the outer asteroid belt indicates that the putative secular resonance sweeping was confined to the inner asteroid belt. 

Several theoretical studies of the dynamical excitation of the asteroid belt linked to the orbital migration history of the giant planets have been carried out in recent years, but these have made use of observational constraints from only the small sample of large asteroids~\citep[e.g.,][]{Minton:2011,Roig:2015,Izidoro:2015,Morbidelli:2015,Toliou:2016}.  That small sample size allows only rough constraints on theoretical models.  We have shown here that the larger asteroids and the smaller asteroids have statistically similar proper eccentricities. Thus, the much larger sample size of the observationally complete MBAs of $H\leq15$ offers greater statistical confidence and more detailed constraints on theoretical models, at least with regard to dynamical excitation mechanisms.  Moreover, differences between the inner and outer asteroid belt can also be examined with this large sample, which is difficult with the small sample of large asteroids, allowing to test and improve theoretical models.

It is natural to ask if the inclination distribution of the observationally complete set of MBAs shares the trends found in their eccentricity distribution.  We are investigating this question, and find that the inclination distribution as well as inclination-eccentricity correlations merit a separate report. 

\section*{Acknowledgements}
This research was supported by NSF (grant AST-1312498), and made use of the NASA Astrophysics Data System Bibliographic Services and the Asteroids Dynamic data service for minor planet data.  We thank Hilke Schlichting, Kathryn Volk and Andrew Youdin for helpful discussions, and Stanley F.~Dermott for comments on the manuscript.

\bibliographystyle{mnras}

\bsp	
\label{lastpage}
\end{document}